\documentclass{article}%
\usepackage{amsfonts}
\usepackage{amsmath}
\usepackage{hyperref}
\usepackage{amssymb}
\usepackage{graphicx}%
\begin{document}

\title{Potentials versus Geometry\thanks{Not quite the same as
\textit{\href{https://en.wikipedia.org/wiki/Ford_v_Ferrari}{Ford v Ferrari}}.
Or is it?}}
\author{T. Curtright$^{\S }$ and S. Subedi$^{\infty}\medskip$\\Department of Physics, University of Miami, Coral Gables, Florida 33124\\$^{\S }${\footnotesize curtright@miami.edu\ \ \ \ \ }$^{\infty}$%
{\footnotesize sushil.subedi04@gmail.com}}
\maketitle

\begin{abstract}
We discuss some equivalence relations between the non-relativistic quantum
mechanics for particles subjected to potentials and for particles moving
freely on background geometries. \ In particular, we illustrate how selected
geometries can be used to regularize singular potentials.

\end{abstract}

Given a solution to Einstein's theory of gravity acting as a background
spacetime geometry, it is well-known that particle motion on this fixed
geometry can be described by an effective potential \cite{MTW}. \ But given a
potential, it is perhaps less widely recognized that an equivalent particle
dynamics can be described by an effective geometry. \ This equivalence is
discussed here in the context of non-relativistic quantum mechanics. \ That is
to say, only the non-relativistic form of Schr\"{o}dinger's equation is used
in our discussion, and only spatial dimensions are allowed non-trivial
geometry, while time is taken to be common to all frames and universal to all observers.

Potential problems can be related to geometrical models in ways that often
make the physics easier to extract and understand from the perspective of one
or the other side of the relationship. \ For a generic potential the
corresponding geometry can be singular, especially if the potential is
singular (e.g. inverse powers of $r$). \ On the other hand, non-singular
geometries can give regularized forms of singular potentials that specify the
essential physical features of those potentials without mathematical
ambiguities or pathologies.

Consider a non-relativistic particle of mass $\mu$\ with energy $E=\hbar
^{2}\varepsilon/\left(  2\mu\right)  $, subject to a potential $V\left(
\overrightarrow{r}\right)  =\hbar^{2}U\left(  \overrightarrow{r}\right)
/\left(  2\mu\right)  $, and whose wave function is a solution of the
Schr\"{o}dinger equation in flat Euclidean space.
\begin{equation}
-\nabla^{2}\Psi+U\Psi=\varepsilon\Psi\label{PotentialSE}%
\end{equation}
Equivalence between this potential problem and a suitably chosen non-flat
geometry can be obtained by relating the solutions of (\ref{PotentialSE}) with
those for \emph{free} particle motion on a specified manifold, written as
\begin{equation}
-\frac{1}{\sqrt{g}}\partial_{\mu}\left(  \sqrt{g}g^{\mu\nu}\partial_{\nu}%
\Psi\right)  =\varepsilon\Psi\label{GeometricSE}%
\end{equation}
In particular, given a rotationally invariant potential, $U\left(  r\right)
$, for a particle with specified angular momentum there is an equivalent
radial Schr\"{o}dinger equation that describes a particle moving freely on a
curved space whose geometry is nontrivial, and vice versa. \ The main result
can be expressed in general as an integro-differential modification of
\href{https://en.wikipedia.org/wiki/Riccati_equation}{the Riccati
equation}\ \cite{Ince}.

In the case of two spatial dimensions (2D) with Euclidean signature, for a
particle with angular momentum $m$ so that the particle's wave function
factorizes as $\Psi\left(  r,\theta\right)  =\Psi_{m}\left(  r\right)
\exp\left(  im\theta\right)  $,\ the main result is
\begin{equation}
\frac{dW}{dr}+W^{2}+\frac{m^{2}}{R_{0}^{2}}\exp\left(  -4\int_{r_{0}}%
^{r}W\left(  \mathfrak{r}\right)  d\mathfrak{r}\right)  =U\left(  r\right)
+\frac{m^{2}-1/4}{r^{2}} \label{2DGeomFromPotl}%
\end{equation}
Upon solving (\ref{2DGeomFromPotl}), the function $W$ encodes the geometry in
terms of the invariant distance on the curved 2D space as%
\begin{equation}
\left(  ds\right)  ^{2}=\left(  dr\right)  ^{2}+R^{2}\left(  r\right)  \left(
d\theta\right)  ^{2}\text{ \ \ with \ \ }R\left(  r\right)  =R_{0}\exp\left(
2\int_{r_{0}}^{r}W\left(  \mathfrak{r}\right)  d\mathfrak{r}\right)  \text{
\ \ i.e. \ \ }W\left(  r\right)  =\frac{d}{dr}\ln\sqrt{R\left(  r\right)  }
\label{2DMetric}%
\end{equation}
and the invariant Laplacian on the 2-manifold is%
\begin{equation}
\frac{1}{\sqrt{g}}\partial_{\mu}\left(  \sqrt{g}g^{\mu\nu}\partial_{\nu
}\right)  =\frac{1}{R\left(  r\right)  }\partial_{r}\left(  R\left(  r\right)
\partial_{r}\right)  +\frac{1}{R\left(  r\right)  ^{2}}\partial_{\theta}%
^{2}=\partial_{r}^{2}+2W\left(  r\right)  \partial_{r}+\frac{1}{R\left(
r\right)  ^{2}}\partial_{\theta}^{2} \label{2DInvLaplacian}%
\end{equation}
Alternatively, given the geometry of the surface with a specified $R\left(
r\right)  $, the corresponding effective potential in flat space for angular
momentum $m$ is given by%
\begin{equation}
U\left(  r\right)  =\frac{1}{2}\frac{R^{\prime\prime}\left(  r\right)
}{R\left(  r\right)  }-\frac{1}{4}\left(  \frac{R^{\prime}\left(  r\right)
}{R\left(  r\right)  }\right)  ^{2}+\left(  \frac{m}{R\left(  r\right)
}\right)  ^{2}-\frac{m^{2}-\frac{1}{4}}{r^{2}} \label{2DPotlFromGeom}%
\end{equation}
As should be expected, for $R\left(  r\right)  \neq r$ this effective radial
potential will depend on the angular momentum.

To establish the relation between the potential problem in 2D flat space and
the 2D curved space system without a potential, it is only necessary to
eliminate the first derivative term in (\ref{2DInvLaplacian}) by writing
$\Psi\left(  r,\theta\right)  =\exp\left(  -\int_{r_{0}}^{r}W\left(
\mathfrak{r}\right)  d\mathfrak{r}\right)  \psi\left(  r,\theta\right)  $ to
obtain%
\begin{equation}
\left(  \partial_{r}^{2}+2W\left(  r\right)  \partial_{r}+\frac{1}{R\left(
r\right)  ^{2}}\partial_{\theta}^{2}\right)  \Psi\left(  r,\theta\right)
=\exp\left(  -\int_{r_{0}}^{r}W\left(  \mathfrak{r}\right)  d\mathfrak{r}%
\right)  \left(  \partial_{r}^{2}+\frac{1}{R\left(  r\right)  ^{2}}%
\partial_{\theta}^{2}-W^{\prime}-W^{2}\right)  \psi\left(  r,\theta\right)
\label{GeomRadial}%
\end{equation}
and compare this to the flat space system with a potential $U$, as obtained by
writing $\Psi\left(  r,\theta\right)  =\psi\left(  r,\theta\right)  /\sqrt{r}%
$, namely,
\begin{equation}
\left(  \partial_{r}^{2}+\frac{1}{r}\partial_{r}+\frac{1}{r^{2}}%
\partial_{\theta}^{2}-U\left(  r\right)  \right)  \Psi\left(  r,\theta\right)
=\frac{1}{\sqrt{r}}\left(  \partial_{r}^{2}+\frac{1}{r^{2}}\partial_{\theta
}^{2}-U\left(  r\right)  +\frac{1}{4r^{2}}\right)  \psi\left(  r,\theta
\right)  \label{PotlRadial}%
\end{equation}
Separating variables as $\psi\left(  r,\theta\right)  =\psi_{m}\left(
r\right)  \exp\left(  im\theta\right)  $ for both (\ref{GeomRadial}) and
(\ref{PotlRadial}) leads to the same second-order radial equation provided%
\begin{equation}
-\frac{m^{2}}{R\left(  r\right)  ^{2}}-W^{\prime}-W^{2}=-\frac{\left(
m^{2}-1/4\right)  }{r^{2}}-U\left(  r\right)
\end{equation}
and hence (\ref{2DGeomFromPotl}) with $R\left(  r\right)  =R_{0}\exp\left(
2\int_{r_{0}}^{r}W\left(  \mathfrak{r}\right)  d\mathfrak{r}\right)  $, or
alternatively (\ref{2DPotlFromGeom}).

In the case of $N$ spatial dimensions (ND) with Euclidean signature
\cite{Green}, on a manifold with isotropic metric of the same form as
(\ref{2DMetric}), a particle with angular momentum $\ell$ has a wave function
that factorizes as $\Psi=\Psi_{\ell}\left(  r\right)  Y_{lm_{1}m_{2}\cdots
m_{N-2}}\left(  \Omega\right)  $,\ with all the angular dependence in the
\href{https://en.wikipedia.org/wiki/Spherical_harmonics#Higher_dimensions}{hyperspherical
harmonics} $Y_{lm_{1}m_{2}\cdots m_{N-2}}\left(  \Omega\right)  $. \ The
invariant Laplacian acting on $\Psi$ is
\begin{equation}
\frac{1}{\sqrt{g}}\partial_{\mu}\left(  \sqrt{g}g^{\mu\nu}\partial_{\nu
}\right)  \Psi=\frac{1}{R\left(  r\right)  ^{N-1}}~\partial_{r}\left(
R\left(  r\right)  ^{N-1}\partial_{r}\right)  \Psi-\frac{1}{R\left(  r\right)
^{2}}~L^{2}\Psi
\end{equation}
where all the $N-1$ angular derivatives are contained in $L_{jk}%
\equiv-i\left(  x_{j}\partial_{k}-x_{k}\partial_{j}\right)  $ with
$L^{2}\equiv\sum_{1\leq j<k\leq N}L_{jk}L_{jk}$. \ The $Y_{lm_{1}m_{2}\cdots
m_{N-2}}$ are eigenfunctions of $L^{2}$ and form a complete set on the
hypersphere, $S_{N-1}$, analogous to the familiar
\href{https://en.wikipedia.org/wiki/Spherical_harmonics}{spherical harmonics}
$Y_{lm}$ on $S_{2}$. \ Acting on $Y_{lm_{1}m_{2}\cdots m_{N-2}}$ the $L^{2}%
\ $eigenvalues are given by \cite{Sommerfeld}
\begin{equation}
L^{2}Y_{lm_{1}m_{2}\cdots m_{N-2}}=l\left(  l+N-2\right)  Y_{lm_{1}m_{2}\cdots
m_{N-2}}%
\end{equation}
for $l=0,1,2,\cdots$, generalizing the well-known $N=3$ case. \ The eigenvalue
equation (\ref{GeometricSE}) then reduces to the radial equation%
\begin{equation}
\frac{-1}{R\left(  r\right)  ^{N-1}}~\partial_{r}\left(  R\left(  r\right)
^{N-1}\partial_{r}\Psi_{l}\left(  r\right)  \right)  +\frac{l\left(
l+N-2\right)  }{R\left(  r\right)  ^{2}}\Psi_{l}\left(  r\right)
=\varepsilon\Psi_{l}\left(  r\right)  \text{ \ \ for \ \ }0\leq r\leq\infty
\end{equation}
Comparison to a potential system in ND flat space, (\ref{PotentialSE}), again
leads to the main result in the form of a modified Riccati equation.
\begin{equation}
\frac{dW}{dr}+W^{2}+\frac{\ell\left(  \ell+N-2\right)  }{R_{0}^{2}}%
~\exp\left(  \frac{4}{1-N}\int_{r_{0}}^{r}W\left(  \mathfrak{r}\right)
d\mathfrak{r}\right)  =U\left(  r\right)  +\frac{\left(  \ell+\frac{1}%
{2}\left(  N-1\right)  \right)  \left(  \ell+\frac{1}{2}\left(  N-3\right)
\right)  }{r^{2}}\label{NDGeomFromPotl}%
\end{equation}
Upon solving for the function $W$, the effective geometry is encoded in the
invariant distance on the curved ND space as%
\begin{equation}
\left(  ds\right)  ^{2}=\left(  dr\right)  ^{2}+R^{2}\left(  r\right)  \left(
d\Omega\right)  ^{2}\text{ \ \ with \ \ }R\left(  r\right)  =R_{0}\exp\left(
\frac{2}{N-1}\int_{r_{0}}^{r}W\left(  \mathfrak{r}\right)  d\mathfrak{r}%
\right)  \text{ \ \ i.e. \ \ }W\left(  r\right)  =\frac{N-1}{2}\frac{d}{dr}\ln
R\left(  r\right)  \label{NDMetric}%
\end{equation}
Alternatively, given the geometry of the surface with a specified $R\left(
r\right)  $, the corresponding potential in flat space for angular momentum
$\ell$ is given by%
\begin{equation}
U\left(  r\right)  =\frac{\left(  N-1\right)  }{2}\frac{d^{2}}{dr^{2}}\ln
R\left(  r\right)  +\left(  \frac{N-1}{2}\right)  ^{2}\left(  \frac{d}{dr}\ln
R\left(  r\right)  \right)  ^{2}+\frac{\ell\left(  \ell+N-2\right)  }%
{R^{2}\left(  r\right)  }-\frac{\left(  \ell+\frac{1}{2}\left(  N-1\right)
\right)  \left(  \ell+\frac{1}{2}\left(  N-3\right)  \right)  }{r^{2}%
}\label{NDPotlFromGeom}%
\end{equation}
As in the 2D case, for $R\left(  r\right)  \neq r$ this effective radial
potential will depend on the angular momentum.

For $\ell=0$ \textquotedblleft s-wave\textquotedblright\ solutions,
(\ref{NDGeomFromPotl}) is an \emph{unmodified} Riccati equation, a first-order
differential equation whose solutions are well-studied \cite{Ince}. \ For a
$1/r$ potential in ND, \emph{plus a constant}, the $\ell=0$ case alone
provides an illustration of potential $\Longrightarrow$\ effective geometry.
\ Let
\begin{equation}
U\left(  r\right)  =\frac{\kappa}{r}+\frac{\kappa^{2}}{\left(  N-1\right)
^{2}}%
\end{equation}
so that (\ref{NDGeomFromPotl})\ for $\ell=0$ becomes%
\begin{equation}
\frac{dW}{dr}+W^{2}=\frac{\kappa}{r}+\frac{\kappa^{2}}{\left(  N-1\right)
^{2}}+\frac{\left(  N-1\right)  \left(  N-3\right)  }{4r^{2}}%
\end{equation}
A particularly simple, rational solution is immediately seen to be%
\begin{equation}
W\left(  r\right)  =\frac{N-1}{2r}+\frac{\kappa}{N-1} \label{particular 1/r}%
\end{equation}
with general solutions obtained by quadrature \cite{Ince}. \ 

The 3D case is exceptional in that there is no $1/r^{2}$ term on the RHS of
the Riccati equation. \ This physically interesting case readily admits
another solution for a pure Coulomb potential, sans constant,
namely,\footnote{NB \ This is the $1/r$ potential expressed as it would be in
\href{https://en.wikipedia.org/wiki/Supersymmetric_quantum_mechanics}{supersymmetric
quantum mechanics}.}%
\begin{gather}
\frac{dW\left(  r\right)  }{dr}+W^{2}\left(  r\right)  =\frac{\kappa}{r}\text{
\ \ for \ \ }N=3\text{ \ \ with}\nonumber\\
W\left(  r\right)  =\sqrt{\frac{\kappa}{r}}\frac{I_{0}\left(  2\sqrt{\kappa
r}\right)  }{I_{1}\left(  2\sqrt{\kappa r}\right)  }=\frac{1}{r}+\frac{\kappa
}{2}-\frac{\kappa^{2}r}{12}+O\left(  r^{2}\right)
\end{gather}
which is well-approximated by (\ref{particular 1/r}) for small $r$. \ But as
before, this is a solution only for $\ell=0$.

The effective geometry corresponding to the solution (\ref{particular 1/r}) is
given by%
\begin{equation}
R\left(  r\right)  =Kre^{2\kappa r/\left(  N-1\right)  ^{2}}\ ,\ \ \ K=\frac
{R_{0}}{r_{0}}~e^{-2\kappa r_{0}/\left(  N-1\right)  ^{2}}%
\end{equation}
The invariant distance for this particular ND manifold is then%
\begin{equation}
\left(  ds\right)  ^{2}=\left(  dr\right)  ^{2}+K^{2}r^{2}e^{4\kappa r/\left(
N-1\right)  ^{2}}\left(  d\Omega\right)  ^{2}%
\end{equation}
In terms of more conventional isotropic coordinates, such that $\left(
ds\right)  ^{2}=\left(  \frac{dr\left(  \rho\right)  }{d\rho}\right)
^{2}\left(  d\rho\right)  ^{2}+\rho^{2}\left(  d\Omega\right)  ^{2}$, let%
\begin{equation}
\rho^{2}=K^{2}r^{2}e^{4\kappa r/\left(  N-1\right)  ^{2}}%
\end{equation}
and solve for $r\left(  \rho\right)  $ in terms of the
\href{https://en.wikipedia.org/wiki/Lambert_W_function}{Lambert W function}.
\ For a repulsive potential $\kappa\geq0$, and thus%
\begin{equation}
r\left(  \rho\right)  =\frac{\left(  N-1\right)  ^{2}}{2\kappa}%
~\operatorname{LambertW}\left(  \frac{2\kappa\rho}{K\left(  N-1\right)  ^{2}%
}\right)  \ ,\ \ \ \frac{dr\left(  \rho\right)  }{d\rho}=\frac{\left(
N-1\right)  ^{2}}{2\kappa\rho}\left(  \frac{\operatorname{LambertW}\left(
\frac{2\kappa\rho}{K\left(  N-1\right)  ^{2}}\right)  }%
{1+\operatorname{LambertW}\left(  \frac{2\kappa\rho}{K\left(  N-1\right)
^{2}}\right)  }\right)
\end{equation}
The resulting geometry is described by%
\begin{equation}
\left(  ds\right)  ^{2}=\left(  \frac{\left(  N-1\right)  ^{2}}{2\kappa\rho
}\frac{\operatorname{LambertW}\left(  \frac{2\kappa\rho}{K\left(  N-1\right)
^{2}}\right)  }{1+\operatorname{LambertW}\left(  \frac{2\kappa\rho}{K\left(
N-1\right)  ^{2}}\right)  }\right)  ^{2}\left(  d\rho\right)  ^{2}+\rho
^{2}\left(  d\Omega\right)  ^{2}%
\end{equation}
A canonical embedding of this $N$-manifold into an $N+1$ Lorentz space is
given by%
\begin{equation}
\left(  ds\right)  ^{2}=\left(  d\rho\right)  ^{2}+\rho^{2}\left(
d\Omega\right)  ^{2}-c^{2}\left(  dt\left(  \rho\right)  \right)  ^{2}%
\end{equation}
with%
\begin{equation}
c\frac{dt\left(  \rho\right)  }{d\rho}=\frac{\sqrt{\left(
1+\operatorname{LambertW}\left(  \frac{2\kappa\rho}{K\left(  N-1\right)  ^{2}%
}\right)  \right)  ^{2}-\left(  \frac{\left(  N-1\right)  ^{2}}{2\kappa\rho
}\operatorname{LambertW}\left(  \frac{2\kappa\rho}{K\left(  N-1\right)  ^{2}%
}\right)  \right)  ^{2}}}{1+\operatorname{LambertW}\left(  \frac{2\kappa\rho
}{K\left(  N-1\right)  ^{2}}\right)  } \label{TimeEmbedding}%
\end{equation}
For $\kappa\geq0$ this is a real embedding all the way down to $\rho=0$ if
$K\geq1$. \ For $K<1$ the embedding is real only for $\rho_{\min}\leq\rho
\leq\infty$ where $\rho_{\min}$ is given by the positive real root of the
radicand in (\ref{TimeEmbedding}).

Graphical representations of the embedded surface and generalizations to
situations where $\ell\neq0$ are left as an exercise for the reader, as is the
$\kappa<0$ situation. \ But it should already be evident from the $\ell=0$
case that a repulsive $1/r$ model is more easily understood as a potential
problem than it is from the geometrical side of the relationship. \ Rather
than pursue the quantum mechanics on the resulting ND manifold for this
example, we consider next a more troublesome singular potential which can be
regularized by mapping onto a well-known geometry.

Consider a smooth, spatial \textquotedblleft bridge\textquotedblright%
\ manifold \cite{ER} (i.e. a \textquotedblleft wormhole\textquotedblright%
\ \cite{Thorne}) given by%
\begin{equation}
\left(  ds\right)  ^{2}=\left(  dw\right)  ^{2}+R^{2}\left(  w\right)  \left(
d\Omega\right)  ^{2}\ ,\ \ \ -\infty\leq w\leq+\infty\label{WormholeMetric}%
\end{equation}
where $R\left(  w\right)  >0$ has an absolute minimum at $w=0$, and behaves
asymptotically as $R^{2}\left(  w\right)  \underset{w\rightarrow\pm
\infty}{\sim}w^{2}+O\left(  1/w^{2}\right)  $. \ For example, the $w$-form of
a static Ellis metric \cite{Ellis} in $N$ spatial dimensions is simply
\begin{equation}
R^{2}\left(  w\right)  =R_{0}^{2}+w^{2}\ ,\ \ \ -\infty\leq w\leq
+\infty\label{Ellis}%
\end{equation}
The metric (\ref{WormholeMetric}) leads to the invariant Laplacian
\begin{equation}
\frac{1}{\sqrt{g}}\partial_{\mu}\left(  \sqrt{g}g^{\mu\nu}\partial_{\nu}%
\Psi\right)  =\frac{1}{R\left(  w\right)  ^{N-1}}~\partial_{w}\left(  R\left(
w\right)  ^{N-1}\partial_{w}\right)  -\frac{1}{R\left(  w\right)  ^{2}}~L^{2}%
\end{equation}
where again all the $N-1$ angular derivatives are contained in $L^{2}$. \ 

The non-relativistic Schr\"{o}dinger energy eigenvalue problem for a particle
moving freely on this manifold is again solved by separating variables.
\begin{equation}
\frac{1}{\sqrt{g}}\partial_{\mu}\left(  \sqrt{g}g^{\mu\nu}\partial_{\nu}%
\Psi\right)  +\varepsilon\Psi=0\ ,\ \ \ \Psi=\Psi_{l}\left(  w\right)
Y_{lm_{1}m_{2}\cdots m_{N-2}}\left(  \Omega\right)
\end{equation}
The effective radial equation is now%
\begin{equation}
\frac{1}{R\left(  w\right)  ^{N-1}}~\partial_{w}\left(  R\left(  w\right)
^{N-1}\partial_{w}\Psi_{l}\left(  w\right)  \right)  +\left(  \varepsilon
-\frac{l\left(  l+N-2\right)  }{R\left(  w\right)  ^{2}}\right)  \Psi
_{l}\left(  w\right)  =0 \label{RadialWormhole}%
\end{equation}
Rather than compare this to rotationally invariant potential scattering in ND
flat space, as above, instead compare this eigenvalue equation to
one-dimensional potential scattering on the line $-\infty\leq w\leq+\infty$,
as given by%
\begin{equation}
\partial_{w}^{2}\Psi_{l}\left(  w\right)  +\left(  \varepsilon-U\left(
w\right)  \right)  \Psi_{l}\left(  w\right)  =0 \label{OnALine}%
\end{equation}
To establish the sought-for relation, again just eliminate the first
derivative term in (\ref{RadialWormhole}) by writing $\Psi_{l}\left(
w\right)  =\left(  R\left(  w\right)  \right)  ^{\left(  1-N\right)  /2}%
\psi_{l}\left(  w\right)  $ to obtain%
\begin{align}
&  \left(  \partial_{w}^{2}+\left(  N-1\right)  \frac{R^{\prime}}{R}%
\partial_{w}+\left(  \varepsilon-\frac{l\left(  l+N-2\right)  }{R\left(
w\right)  ^{2}}\right)  \right)  \Psi_{l}\left(  w\right) \nonumber\\
&  =R^{p}\left(  w\right)  \left(  \partial_{w}^{2}+\left(  \varepsilon
-\frac{1}{2}\left(  N-1\right)  \frac{R^{\prime\prime}}{R}-\frac{1}{4}\left(
N-1\right)  \left(  N-3\right)  \left(  \frac{R^{\prime}}{R}\right)
^{2}-\frac{l\left(  l+N-2\right)  }{R^{2}}\right)  \right)  \psi_{l}\left(
w\right)  \label{2ndOrder}%
\end{align}
where primes indicate derivatives with respect to $w$. \ 

The second-order equation (\ref{2ndOrder}) is the same as that for potential
scattering on the line (\ref{OnALine}) if
\begin{equation}
U\left(  w\right)  =\frac{1}{2}\left(  N-1\right)  \frac{R^{\prime\prime}}%
{R}+\frac{1}{4}\left(  N-1\right)  \left(  N-3\right)  \left(  \frac
{R^{\prime}}{R}\right)  ^{2}+\frac{l\left(  l+N-2\right)  }{R^{2}%
}\label{LinePotl}%
\end{equation}
For the Ellis metric (\ref{Ellis}), $R^{\prime}=w/R$, $R^{\prime\prime}%
=R_{0}^{2}/R^{3}$, so%
\begin{equation}
U\left(  w\right)  =\left(  N+2l-3\right)  \left(  N+2l-1\right)  \frac
{1}{4R\left(  w\right)  ^{2}}-\left(  N-1\right)  \left(  N-5\right)
\frac{R_{0}^{2}}{4R\left(  w\right)  ^{4}}\label{NDEllisPotl}%
\end{equation}
In particular, for 3D,%
\begin{equation}
U\left(  w\right)  =\frac{l\left(  l+1\right)  }{R_{0}^{2}+w^{2}}+\frac
{R_{0}^{2}}{\left(  R_{0}^{2}+w^{2}\right)  ^{2}}\text{ \ \ for \ \ }%
N=3\label{3DEllisPotl}%
\end{equation}
This $U\left(  w\right)  $ may be thought of as a regularized form of
$1/r^{4}$, a singular potential that has been well-studied in 3D
\cite{Singular1950,Singular1971}.

The angular momentum \emph{in}dependent $1/R^{4}$ term in the potential
(\ref{NDEllisPotl}) is repulsive for $N\leq4$, absent for $N=5$, and
attractive for $N\geq6$. \ (It so happens to be a regularized form of an
attractive electrostatic potential for a point particle for $N=6$
\cite{Green,C et al.}.) \ In any case, non-relativistic scattering for the
regular potential $U\left(  w\right)  $ has both elastic and inelastic
components for all $N\geq2 $, where elastic scattering is interpreted as both
incident and emergent probability flux on the \textquotedblleft
upper\textquotedblright\ branch of the wormhole manifold, and inelastic
scattering is to be understood as emergent flux on the \textquotedblleft
lower\textquotedblright\ branch of the manifold, without any incident flux on
that lower branch. \ For incident flux on the wormhole's upper branch, with
$R_{0}>0$, there is always some probability flow through the bridge joining
the two branches so that particles are effectively absorbed by the wormhole,
as viewed by an observer located on the upper branch. \ For any $N$, the time
independent partial wave scattering amplitudes can be determined exactly for
$U\left(  w\right)  $ in terms of spheroidal wave functions \cite{DLMF}. \ An
Argand diagram of the phase shifts clearly reveals the inelastic behavior.
\ This is discussed in considerable detail in \cite{CS}.

The singular $1/r^{4}$ potential that arises from $U\left(  w\right)  $ in the
limit as $R_{0}\rightarrow0$ is somewhat problematic, as explained in
\cite{Singular1971}\ and references cited therein. \ At issue is whether the
Hamiltonian for such a potential admits a
\href{https://en.wikipedia.org/wiki/Self-adjoint_operator#:~:text=Self-adjoint%20operators%20are%20used%20in%20functional%20analysis%20and,represented%20by%20self-adjoint%20operators%20on%20a%20Hilbert%20space.}{self-adjoint}
extension for $0\leq r\leq\infty$, with only real eigenvalues and unitary time
evolution. \ For a repulsive $1/r^{4}$ this can be arranged with an
appropriate choice of boundary conditions at $r=0$. \ An attractive $1/r^{4}%
$\ is another story, however, with considerable ambiguity in the boundary
conditions at the origin. \ As stated in \cite{Singular1971}:\bigskip

\begin{quote}
\textquotedblleft Thus the basic feature of an attractive singular potential
is seen to lie in the fact that physical processes are not uniquely
determined. \ This gives rise to the possibility of imposing unusual or
unconventional boundary conditions in physical problems as a means of
representing particular physical processes. \ An example of a process of this
type is provided by particle absorption or capture.\textquotedblright\bigskip

\textquotedblleft Since the deficiency indices are infinite, nonself-adjoint
extensions are possible which would correspond to boundary conditions which
describe inelastic scattering.\textquotedblright\bigskip

\textquotedblleft Thus, in the singular case, the long-range part of the force
between particles does not alone suffice to determine their behavior; some
cutoff mechanism apparently must be provided.\textquotedblright\bigskip
\end{quote}

The regularization defined by the wormhole geometry avoids all these
ambiguities. \ The Hamiltonian for a non-relativistic system governed by
$U\left(  w\right)  $ is manifestly self-adjoint for $R_{0}>0$ when defined on
a \href{https://en.wikipedia.org/wiki/Rigged_Hilbert_space}{rigged Hilbert
space} with the usual boundary conditions as $w\rightarrow\pm\infty$. \ But
there is a price to be paid for this mathematical convenience, namely,
inelastic scattering, even when the potential is repulsive. \ This feature for
$U\left(  w\right)  $ is discussed further in \cite{CS}.

In conclusion, the paper has exhibited various relationships between
non-relativistic quantum systems involving a potential, in flat space, and
systems without a potential but defined on curved manifolds. \ The main
general results are encoded in (\ref{NDGeomFromPotl}), (\ref{NDMetric}),
(\ref{NDPotlFromGeom}), and (\ref{LinePotl}) for $N$-dimensional spatial
manifolds. \ More specific examples have been presented involving $1/r$
potentials and regularized $1/R^{4}$ potentials on wormhole manifolds.
\ Perhaps our discussion should be viewed as a contemporary reconsideration of
Riemann's ideas\ about dynamics in terms of geometry \cite{Riemann}. \ Or
perhaps not. \ The reader can decide.

\end{document}